\documentclass{aastex}

\begin{document}
\title{A Search for CO Absorption in the Transmission Spectrum of HD~209458~b}
\author{Timothy M. Brown\altaffilmark{1}, Kenneth G. Libbrecht\altaffilmark{3},
David Charbonneau\altaffilmark{2,1}}
\altaffiltext{1}{High Altitude Observatory/National Center for Atmospheric
Research, 3450 Mitchell Lane, Boulder, CO 80307.
The National Center for Atmospheric Research is sponsored
by the National Science Foundation.}
\email{timbrown@hao.ucar.edu}
\altaffiltext{2}{Harvard-Smithsonian Center for Astrophysics,
60 Garden St., Cambridge, MA 02138}
\email{dc@caltech.edu}
\altaffiltext{3}{California Institute of Technology, 1200 E. California Blvd.,
Pasadena, CA 91125}
\email{kgl@caltech.edu}

\begin{abstract}
We observed one transit of the extrasolar planet HD 209458b
with the NIRSPEC spectrograph on the Keck II telescope.
Using time series of low-noise observations in the wavelength
range 2.0 $\mu$m to 2.5 $\mu$m, we searched for extra absorption
from the 1st overtone rotation-vibration band of CO near
2.3 $\mu$m.
This was not detected, with a detection limit that fails to test
simple models of the planetary atmosphere by a factor of about 3.
Great improvements in the detectability of the CO spectrum features
could be realized by observing a transit that is centered near
stellar meridian passage, and in better weather.
Since it appears that similar observations taken under better
circumstances might succeed,
we describe our analysis procedures in detail. 
\end{abstract}


\section{Introduction}

Since the discovery of the first planet 
of a distant Sun-like star \citep{may95},
radial velocity measurements have succeeded in locating about
80 such planets \citep[see the review by][]{ppiv}.
Although radial velocity (RV) techniques are well suited to finding
such planets, they convey little information about the planets'
physical nature; the inferences from such data are limited
to the period and eccentricity of the planetary orbit and
a lower limit on the planet's mass.
To date, the only effective probes of the structure or composition
of an extrasolar planet involve observations taken during
transits, when the planet passes across the face of its parent star.
And to date, only the planet circling HD 209458 is known to show
such transits \citep{hen00,cha00}.

Accurate photometry of transits yields estimates of the planetary
radius; together with a mass estimate from RV data, these give a
density estimate, and hence some information about the planet's
composition and history \citep{maze00,bur00,bro01}.
To learn about the composition in other than a global sense, however,
one must turn to spectroscopy.
During a transit, light that passes from the star through the outer
parts of the planet's atmosphere has impressed on it a spectrographic
signature of the atmospheric constituents \citep{sea00,hub01}
and \citet{bro01a}, henceforth referred to as Br01.

Giant planets in small orbits should have extended atmospheres, because
of their high temperature, predominantly light-gas composition, and
relatively low surface gravity.
Thus, the cross-section for these atmospheres to alter the passing
starlight is comparatively large.
Early attempts to detect the planetary signature in 51 Peg \citep{cou98,
rau00} and in HD 209458 \citep{bun00, mou01} have been unsuccessful.
At least one visible effect,
absorption by atomic sodium in the atmosphere of HD 209458b, 
has however been measured using 
the Hubble Space Telescope \citep{cha02}.

The total absorption from sodium corresponds to a relative intensity
diminution of about 2.4 $\times$ 10$^{-4}$ averaged over a bandpass of
about 1 nm.
Sodium, though spectroscopically very active, is a trace constituent of
the planetary atmosphere.
Some molecules, notably H$_2$O, CO, and CH$_4$, are expected to be much
more abundant species, with greater diagnostic potential.
Among these molecules, CO is of particular interest.
Simple considerations of radiative equilibrium suggest that the
observable part of the atmosphere of HD 209458b should have a temperature
between 1000 K and 1400 K;  Rayleigh scattering from H$_2$ sets a limit of
about 1 bar to the deepest level that can be reached by a tangential light ray.
At such high temperatures and low pressures, CO is expected to be the
dominant carbon-bearing compound, abundant enough that even its relatively
weak transitions (in particular the first overtone rotation-vibration band
redward of 2.3 $\mu$m wavelength) will generate absorption with significant
equivalent width.
Indeed, models (e.g. Br01) show maximum depths for the lines in this
band of up to 10$^{-3}$ relative to the stellar continuum, 
with variations in the total strength of the band
and of the relative strengths of its component lines providing diagnostics
for cloud height, heavy element abundance, and temperature.
Moreover, CO is not present with high abundance in the terrestrial atmosphere,
minimizing the difficulty of separating the planet's signal from
terrestrial contamination.
On the other hand, the 2.3 $\mu$m CO band does lie in a region of significant
CH$_4$ absorption.
Because of methane's high opacity, even the small amount of CH$_4$ in the Earth's
atmosphere is enough to cause problems with the analysis.

The orbital period of HD 209458b is about 3.52 days.
At a given longitude, observable transits occur once per week for
intervals of about two months, while the time of transit occurs
about 70 minutes later in the night for
each passing week.
Since the duration of a transit is 3 hours, there are only a few optimally
timed transits in each observing season.
We describe here an exploratory observation of one transit of this planet
using the NIRSPEC echelle spectrograph. 

\section{Observations}

We observed HD 209458 with the Keck II telescope
and the NIRSPEC echelle spectrograph, the latter operating
in high-resolution (echelle) mode \citep{mac98}.
The transit we observed occurred on UT 2000 Aug. 23;
on this night the NIRSPEC was connected through the Keck II Adaptive
Optics (AO) system.
The spectrograph setup placed 7 orders 
(echelle orders 31 (redmost) to 37 (bluemost))
on the Aladdin 1K x 1K detector,
covering about half of the wavelength range from 2.04 $\mu$m to 2.47 $\mu$m
at a spectral resolution $R\ = \ \lambda/\delta\lambda \ = \ 25000$.
At this resolution, 1 pixel on the detector corresponds to a Doppler shift
of about 4.3 km s$^{-1}$.

We considered alternating observations of HD 209458 with ones of a
reference star of similar spectral type.
Doing so would have provided a control against which to judge the reality
of a detection, as well as a measurement of atmospheric extinction that
was, in some sense, independent of that from the target star.
On the other hand, the anticipated noise level was already large enough
to threaten detectability even with the full observing time devoted to
HD 209548, a few degrees of separation in the sky would be enough to make an
important difference in the extinction, and slewing overhead and repeatability
of positioning on the slit would become issues.
Based on these considerations, we judged that it was better not to observe
a comparison star.

We began taking spectra of HD 209458 at 05:41 UT, 35 minutes after the beginning
of the transit, which occurred at 05:06 UT. 
At the time of the first observation, the star was at an elevation of 
only 25 degrees.
Observations ended at 12:00 UT, when the telescope closed because of fog.
Cirrus reduced transmission during the first two hours of observation,
and observations were interrupted for a short time after the transit's end
because of high humidity.
In all, we obtained 83 spectra of HD 209458, of which 65 were of usable quality.
Of the good spectra, 22 were taken during the transit,
and 43 afterwards.
For all of the spectra, we used an integration time of 200 s.
The in-transit spectra were taken with airmass values between 2.33 and 1.17,
and all of the out-of-transit spectra had airmasses below 1.16.
As we shall elaborate below, this distribution of atmospheric absorption
between the in- and out-of-transit observations causes significant difficulties
in the data analysis.

\section{Data Analysis}

The ultimate objective of the data analysis procedures is to produce
the ratio of averaged spectra taken in and out of transit (see Br01
for details).
In the averaging process, one must take account of the time-varying
Doppler shift of the planet, which changes by $\pm$14 km s$^{-1}$
during the 3-hour duration of the transit.
The analysis therefore proceeded in 4 major steps, as follows.
(1) We corrected the images for known instrumental influences (notably gain and
dark current), and extracted 1-dimensional spectra from the corrected images.
In the process, we obtained estimates of noise, as well as of spectrum shift
and cross-dispersion profile parameters for use in later fits.
(2) By means of regression against the various globally-determined spectrum
parameters,
and by a filtering process based on singular value decomposition (SVD) of
time series of spectra, we removed (as best we could) time-dependent
variations caused by the spectrograph and by the Earth's atmosphere.
(3) We formed an average out-of-transit spectrum, and took the ratio of
each in-transit spectrum to this average.
We cross-correlated these ratio spectra against appropriately Doppler-shifted
model transmission spectra (which in this wavelength region are dominated
by CO absorption), and averaged together the resulting
correlation functions.
A peak in this average correlation would indicate a possible detection
of CO absorption.
(4) To calibrate the analysis process, we injected artificial signals
of known magnitude into the data stream after step (1), and measured
the size of the resulting correlation peak.
This allowed us to estimate the degree to which the complicated fitting
procedures in step (2) attenuate real signals and decrease the
sensitivity of the observations.

\subsection{Calibration and Extraction}

Low-level reduction of the NIRSPEC images started with calibration.
First we corrected each 2-dimensional image for position-dependent
bias and for dark current.
The NIRSPEC detector has very substantial spatial variations in both
its exposure-independent (bias) and its exposure-dependent (dark)
contributions.
We estimated these from suitable median-averages of a dozen or so dark images
taken with long and short integration times.
In addition, many pixels show excess noise, which
may be thought of as time-varying dark current.
We searched for these noisy pixels at two different points in the analysis,
and flagged them so that intensity estimates involving them could be
given low weight when computing the average spectra.
Next we corrected for gain variations by dividing the 2-dimensional images
by flat field images.
These flats were also derived by median-averaging a group of 12 bias- and
dark-corrected wide-slit spectra of the incandescent calibration lamp.
In addition to explicitly flagging noisy pixels, we computed noise estimates
for all pixels, for use in subsequent averaging and fitting operations.

Extraction of 1-dimensional spectra from the 2-dimensional images proceeded
by first estimating the coefficients of a quadratic polynomial describing
the cross-dispersion order position as a function of the along-dispersion
coordinate.
This was done separately for each of the 7 echelle orders, and for each
image.
Motion in the cross-dispersion direction from beginning to end of the night
was about 6 pixels, and sample-to-sample variations in this position were
typically a few tenths of a pixel.
With the order positions known, we estimated the intensity at each wavelength
via an optimally-weighted fit to a cross-dispersion point spread function (PSF).
We estimated the PSF by averaging the intensity parallel to the estimated
order positions for orders 2 through 6 (orders 1 and 7 lying either very
close to or partially outside the detector boundaries).
We assigned weights to each pixel in the PSF fit 
that were inversely proportional 
to the square of their estimated errors, 
including photon noise from the spectrum itself and
bias- and dark-related noise values derived from the calibration process.
The band of pixels included in PSF fitting was 20 pixels wide, and the full
width at half maximum (FWHM) of the orders was typically 4 pixels.
We made no allowance for sky background.
Because of our relatively short exposures combined with the very small
projected slit width produced by the AO system, no sky emission was visible
at any wavelength in our spectra.

After extracting the 1-dimensional spectra, we cleaned each spectrum by
first correcting the values of obviously corrupted pixels, and then by
spatial filtering.
To locate corrupted pixels, we searched for isolated pixels that differed
from the median of their immediate neighbors by more than
5 times their estimated error.
We replaced such points with the local median values.
Typically, only a few points per time sample were 
corrected in this fashion~\--~few enough that
simple replacement by the local median gave an adequate
correction.
Removing them was important, however, in order that the subsequent Fourier
filtering should not introduce large artifacts.
We then spatial-filtered each order individually in the frequency domain,
using a boxcar filter in wavenumber, but modified so that the transition
to zero was through a cosine taper with width equal to 10\% of the total
width of the filter bandpass.
The cutoff frequency was 1 cycle per 3.8 pixels;
this frequency was established by examination of the power spectra of several
low-noise echelle spectra,
and corresponds to the point at which the signal spectrum falls below the
white-noise background in good spectra.
An incidental benefit of filtering the spectra in this manner is to assure that
they are band-limited, and hence that they may be shifted by Fourier
interpolation.

The top trace in Figure 1 shows order 5 (echelle order 33) of 
a well-exposed spectrum 
that has been calibrated, extracted,
and cleaned in the fashion just described.
The signal level is about 2$\times$10$^5$ photoelectrons 
per pixel along the dispersion,
implying a Poisson-statistics-limited signal to noise ratio (SNR) of about 450.
This was near the best signal obtained, but the variability was large;
average spectra had about half this signal level, and some 
(affected by clouds) had much less.
After scaling and shifting along the dispersion 
(usually by a fraction of a pixel),
successive spectra usually agree within errors that are roughly consistent with
the Poisson-statistics SNR.
Over longer time intervals the differences are larger.
Comparison with a model of transmission through the Earth's atmosphere,
shown in the bottom trace in Figure 1, shows that almost all of the absorption
features in the spectrum are telluric, caused mostly by CH$_4$.
At wavelengths below 2.07 $\mu$m one can see absorption from telluric H$_2$O;
the only strong stellar spectrum feature seen
is the hydrogen Br $\gamma$
line at 2.166 $\mu$m.
For use in later regression operations, we estimated the relative
depth of the CH$_4$ features in each time sample from the spectrum in 
order 5 (echelle order 33).

Aside from telluric absorption, the most prominent features of the spectrum are
two sets of interference fringes, 
apparently arising from plane-parallel elements
(most likely windows or filters) in the spectrograph's optics. 
These fringes can be seen most clearly in the top trace in Fig. 1, between
wavelengths of 2.310 and 2.315 $\mu$m.
The two fringe systems have similar amplitudes, each 
modulating the transmission by 
a few percent.
Their wavelength dependences corresponds to optical thicknesses (thickness times
refractive index) of about 1.8 cm and 2.2 cm; the beating between the two fringe
systems (as well as internal structure within the lineshape of each fringe
individually) leads to a rather complicated pattern of transmission.
Moreover, the fringe patterns vary in both phase and amplitude 
on all time scales.
These temporal variations are fairly large, and provide clear evidence
that the fringes arise in the instrument and not, for instance, from
molecular absorption in the stellar spectrum.
We speculate that these variations arise from changes in the effective
incidence angle and f/ number of the light beam striking the fringing
elements, with these changes caused in turn by varying effectiveness of
the adaptive optics correction.
We lack information to verify this hypothesis, however.
As part of the calibration process, we estimated the phases and amplitudes
of each fringe system in order 4 (echelle order 34), which contains few
telluric features.
This 2-sinusoid parameterization gave a poor fit to the actual
fringe pattern, but the phases and amplitudes proved to be useful
external variables in the regression operation that follows.

\subsection{Regression and SVD Filtering}

To remove the remaining variations in atmospheric absorption and in
the fringe pattern, we resorted to a linear regression against
known external variables, followed by a filtering process based
on singular value decomposition of time series of corrected spectra.

In the regression process, we worked not with the observed intensities
$S(x,t)$ as a function of pixel position and time, but rather with
the logarithms of these quantities, denoted by $Q(x,t)$.
This transformation allowed us to express the large multiplicative
changes in intensity (due to clouds, for instance) as additive 
functions, while preserving the form of the much smaller changes
due to other causes.
Thus, we represented the observed spectra as
$$
Q(x,t) \ \equiv \ln S(x,t) \ = \ \left < Q(x) \right > \ + \ 
\sum_j C_j(x) E_j(t) \ + R(x,t) \ \ , \eqno (1)
$$
where $<>$ denotes a time average, the $E_j(t)$ are the estimated time
variation of various external variables, the $C_j(x)$ are
the regression coefficients, determined by minimum-$\chi^2$ fitting
independently at each pixel, and the $R(x,t)$ are the residuals around the
fitted function.

The external variables $E_j(t)$ used in the regression were parameters that
related to the spectrum as a whole, or sometimes to individual spectrum
orders, and that we took to be symptomatic of various perturbations
to the spectrum detection process.
These included the total intensity averaged over each spectrum order,
the relative depth of the telluric CH$_4$ features, the shift
of these features (in pixels) along the dispersion direction,
and the amplitude and phase of the larger-amplitude fringe system,
as measured in order 4.
The depth of telluric features was, more or less, a proxy for airmass.
Better results were obtained using the measured feature depths, however,
so the computed airmass was not included as a parameter in the regression.
As described above, we estimated amplitudes and phases for each of
the 2 fringe systems; we found the corresponding parameters for the
2 systems to be well correlated, however, so we used only one set of
amplitudes and phases in the regression.
Finally, the fringe shifts were large enough ($\pm$ 1 pixel)
compared to the fringe wavelength (about 8 pixels) that it proved
necessary to include as parameters both linear and quadratic functions
of the fringe shift.
After fitting the regression coefficients, we produced corrected spectra
$Q_C(x,t)$ given simply by
$$
Q_C(x,t) \ = \ \left < Q(x) \right > \ + \ R(x,t) \ \ . \eqno (2)
$$
Note that these corrected spectra did not have the effects of fringes
and telluric absorption removed.
Rather, these features were merely adjusted to have their average
values, to the extent that these were consistent with the data
and with our understanding of the observations' systematic errors.

After the regression process, we found that certain systematic time
variations remained in the spectra. 
That is, we found that significant correlations
existed between the time series at different pixels, and that
the responsible time variations did not have the form of any of the
identified external parameters.
To remove these, we started with the matrix of residuals $R(x,t)$
for each echelle order.
We then computed the singular value decomposition of this matrix
\citep{pre86}, which may be written
$$
R(x,t) \ = \ \sum_k^{N} W_k V_k(x) U_k(t) \ \ . \eqno (3)
$$
Here the $N$ vectors $U_k(t)$ are orthonormal and span the space of all 
possible time series of length $N$.
The orthonormal vectors $V_k(x)$ each correspond to one of the $U_k(t)$,
and give the relative amplitude of that time series component at each pixel.
Finally, the scalars $W_k$ (which are ordered from largest to smallest)
give the amplitude of $V_k(x) U_k(t)$ in
the actual time series.
Since the $U_k$ and $V_k$ are both orthonormal, $W_k^2$ describes the
total variance in the time series attributable to $V_k(x) U_k(t)$.
If the $V_k(x)$ corresponding to (say) the largest $W_k$ displays 
components of similar magnitude for many values of $x$, then 
that mode of variation accounts for an unusually large part
of the total variance, and involves strong correlations among pixels.
Such behavior is characteristic of many kinds of instrumental error
(also, alas, of many kinds of signal).
After some experimentation, we concluded that, for these data,
the best compromise
between suppressing noise and attenuating possible real signals
was to truncate the variations corresponding to the 5 largest
singular values.
We thus set $W_1$ to $W_5$ in Eq (3) to zero, and created the filtered
form of $R(x,t)$ by carrying out the indicated matrix multiplications.
The corrected spectra were then formed as in Eq. (2).

\subsection{Spectrum Ratio and Correlation}

To obtain an estimate of the out-of-transit spectrum, we simply averaged
the corrected spectra that we obtained after the end of the planetary
transit, weighting each pixel according to the estimated error for that
wavelength and time sample.
We then formed the ratio of each in-transit spectrum with the averaged
out-of-transit spectrum.

We searched for evidence of CO absorption in the spectrum ratios by
cross-correlating these ratios against a model spectrum that was sampled on the
NIRSPEC wavelength grid.
This approach is similar to that used by \citet{wie01} in their
search for evidence for CH$_4$ in the infrared spectrum of $\tau$ Boo.
We obtained a suitable template spectrum from the fiducial model
of HD 209458b described by Br01.
The model spectrum could not be used directly; first it was necessary
to separate the spectrum of CO from that of other molecules (notably
H$_2$O and CH$_4$). 
As explained in Br01, for species such as CO, which are expected
to be present with a nearly constant mixing ratio throughout the
planetary atmosphere, there is a simple and accurate approximation to the
emerging transmission spectrum.
For such opacity sources, one may write
$$
\Re^\prime(\lambda) \ = \ -R_*^{-2} \left [ R_p^2 \ + \ 2R_p H \ln
({\kappa(\lambda) \over
\kappa_C}) \right ]  \ \ , \eqno (4)
$$
where $1 + \Re^\prime(\lambda)$ is the wavelength-dependent ratio of in- to
out-of-transit intensity,
$R_*$ and $R_p$ are the radii of the star and the planet,
$H$ is the density scale height in the planetary atmosphere,
and $\kappa(\lambda)$ and $\kappa_C$ are the line opacity from
the species under consideration and from the underlying continuum,
respectively.
For this purpose we used estimates of $R_*$ and $R_p$ from 
\citet{bro01}.
We took estimates of $H$ and the opacities from the 0.01-bar level of
the model, and checked the approximation to $\Re^\prime$ against output
from the full model to verify that the line depths were (very nearly) correct.

For each of the spectrum ratio samples calculated above, we formed the
cross-correlation against the model CO spectrum, shifted in wavelength
according to the planet's computed Doppler shift at the time of the measurement.
We corrected the standard spectrum wavelengths for the orbital motion
of the planet, but not for the Earth's orbital motion and rotation
(which cause changes of less than 0.1 pixel during the transit), nor for
the mean radial velocity of the parent star.
A true signal, if present, should therefore have appeared as a correlation
peak at the mean system radial velocity of 
-14.8 km s$^{-1}$ \citep{maze00, nide02}.
The solid line in the top panel of Figure 2 shows the result of
averaging together the cross-correlations for all of the time samples
of order 5,
weighted to account for their calculated errors.
The bottom panel shows the same averaged cross-correlation for order 6.
No significant peak is seen in either correlation function;
clearly the CO transmission spectrum of HD 209458b, if present, is
too weak to be detected in these observations.

\subsection{Calibration of the Detection Algorithm}

The procedure used to calculate the ratio spectra was complicated,
with evident opportunities for attenuating the desired signal in the
process of removing unwanted noise.
We therefore calibrated the detection scheme by injecting simulated
planetary transmission spectra into the calibrated and extracted
in-transit spectra, and reducing these adulterated spectra as we did
the original data.
For this purpose we used the spectrum ratio $\Re^\prime (\lambda)$
computed for the fiducial model described by Br01,
including all opacity sources, not just CO.
We calculated $\Re^\prime$ on a wavelength grid with 2 km s$^{-1}$
resolution, and integrated over NIRSPEC pixels to give the desired
wavelength sampling.
For each observed spectrum sample, we scaled the model spectrum
ratio according to the calculated transit light-curve for that time,
and shifted it in wavelength with the appropriate Doppler shift.
We then simply multiplied the observed spectrum 
by $[1 + f \Re^\prime(\lambda)]$,
where $f$ was an arbitrary scale factor that allowed relatively large
signals to be injected, for easy detection.

We processed the adulterated data just as we did the original spectra.
Results are shown in Figure 2 for $f = 2$ and for $f = 5$.
The size of these correlation peaks, relative to the noise background,
suggests that the current data set is inadequate to reveal the CO
spectrum if it has the same strength as in the model;
for a reliable detection at the nominal band strength, one would need
to improve sensitivity by a factor of about 3.

We performed a similar test in which we injected the artificial signal
after the SVD filtering step, i.e., immediately before computing
the cross-correlations.
In this case, we detected the signal expected from the model ($f = 1$)
with SNR = 3, i.e., with a confidence similar to that attained
in the original test with $f = 3$.
This confirmed our suspicion that the data-cleaning process attenuates
real transit signals.
We subsequently tested several variants of the data cleaning procedure,
but were unable to improve the situation:
techniques that retained more signal amplitude also admitted more
noise, with no gain in the ultimate SNR.
This testing also showed that the loss of signal occurred in the
regression step, i.e., in the
process of fitting out the airmass-dependent variation of atmospheric
absorption, and not in the SVD filtering.

\section{Discussion}

Based on the cross-correlation analysis, we can set a limit to the
CO band strength in HD 209458's in-transit transmission spectrum:
it is no more than about 3 times its strength in Br01's
fiducial model.
From an astrophysical point of view, this limit is not informative.
As indicated in Eq. (4), the strength of the CO absorption feature
is proportional to the logarithm of the line-to-continuum opacity
ratio.
For lines near the CO bandhead, this ratio in the fiducial model
is already about 10$^4$.
Tripling the line strength would require another factor of 10$^{12}$,
which is not possible.

In view of this noise-limited non-detection, one must ask whether
further observations of this sort are worthwhile.
We believe the answer to this question is yes;
it seems likely that significant improvements in detection
sensitivity can be achieved, even with the same facilities as
used in the current study.
First, it should be possible to reduce the Poisson noise in the
observations by a significant factor.
As described above, these observations were heavily compromised by
clouds during the first half of the transit.
Good weather would increase the number of detected in-transit photons
by about a factor of 2.
With good weather throughout, one would also see about twice as large
a range of planetary Doppler shift,
which would help to separate planetary spectrum features from the
telluric CH$_4$ lines.
A further gain in light flux could probably be obtained by using
the system without AO.
With many fewer optical surfaces and a seeing-limited point spread
function (rather than one with a sharp central peak but much light
lost in the far wings, as the AO system produces),
it should be possible to increase system transmission by another
significant factor, perhaps as large as 5.
For observations of such a bright star as HD 209458, the additional
sky background associated with non-AO operation would not cause difficulties.

Even more important than increasing the photon flux, it is possible to
improve the sensitivity by observing a transit that is centered near
the star's meridian crossing.
In the current data set, all of the transit observations occurred
at relatively high airmass, when the terrestrial CH$_4$ features
were stronger by as much as a factor of 2 than during the out-of-transit
observations.
Thus, the time dependence of the terrestrial absorption was qualitatively
the same as that of the sought-after planetary absorption:
large during the transit and small thereafter.
This similarity of functional form caused the data cleaning
procedures to attenuate the transit signal strongly.
By choosing a transit that is more favorably positioned during the night,
this problem can be largely alleviated.

Finally, one can hope for improvements in the NIRSPEC optics that
would remove the fringes that complicate the data analysis.
Even lacking such instrumental improvements, we speculate that
operating without adaptive optics would yield more stable fringes,
since variations in the effective f/ number and incidence angle of
the beam entering the spectrograph would likely be smaller.
The mere presence of fringes is much less damaging than their
temporal variation;
making them more repeatable would greatly improve the prospects
for detecting small time-dependent signals.

\acknowledgements

Data presented herein were obtained at the W.M. Keck Observatory, 
which is operated as a scientific partnership among the California 
Institute of
Technology, the University of California and the 
National Aeronautics and Space Administration. 
The Observatory was made possible by the generous financial
support of the W.M. Keck Foundation.
We wish to recognize and acknowledge the very significant cultural role 
and reverence that the summit of Mauna Kea has always had within the indigenous
Hawaiian community.  
We are most fortunate to have the opportunity to conduct 
observations from this mountain.
We are grateful to Randy Cunningham for his assistance with NIRSPEC
operations,
to Dimitar Sasselov and Sara Seager for many useful conversations,
and to the anonymous referee for helpful suggestions.

\clearpage

\figcaption
{(Top curve) One observed sample of the spectrum of HD 209458,
after calibration, flat-fielding, and spatial filtering.
Shown is order 5 (echelle order 33).
(Bottom curve) Model of the telluric transmission spectrum in
the same wavelength range, displaced vertically for clarity.
All of the features in this part of the spectrum arise from
CH$_4$ in the Earth's atmosphere.
The additional fine-scale ripples seen
in the upper plot are due to fringing in the spectrograph optics.
\label{fig1}}

\figcaption
{(Top panel) Time-averaged cross-correlation of model CO spectrum
against observations, for order 5 (echelle order 33) (solid line).
Also shown are similar cross-correlations using data into which 
have been added artificial signals scaled from the fiducial model
spectrum of Brown (2001).
The scaling factors used were 2 (dashed line) and 5 (dotted line).
(Bottom panel) Same as top panel, but for order 6 (echelle order 32).
\label{fig2}}

\begin{figure}
\plotone{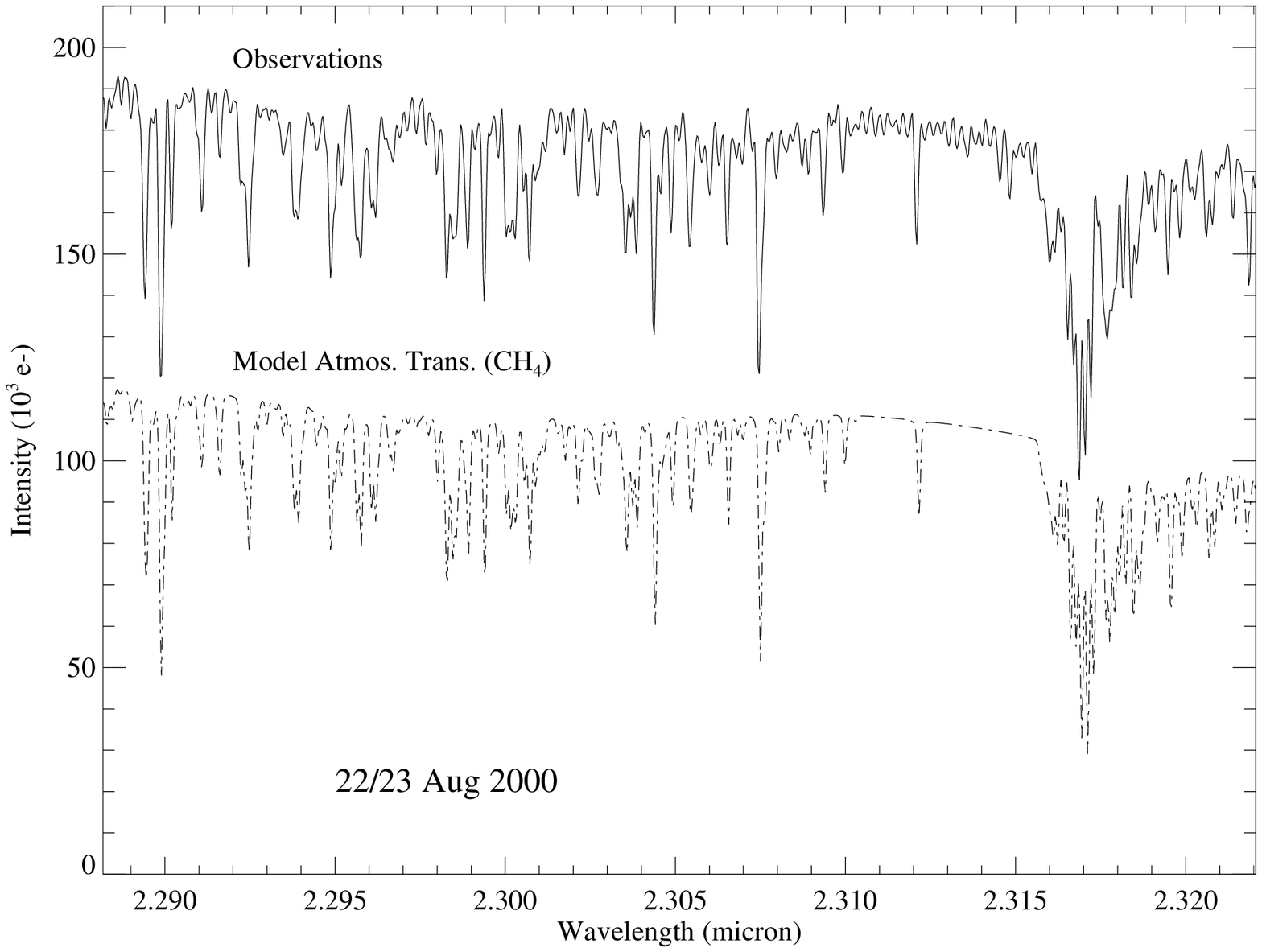}
\end{figure}

\begin{figure}
\plotone{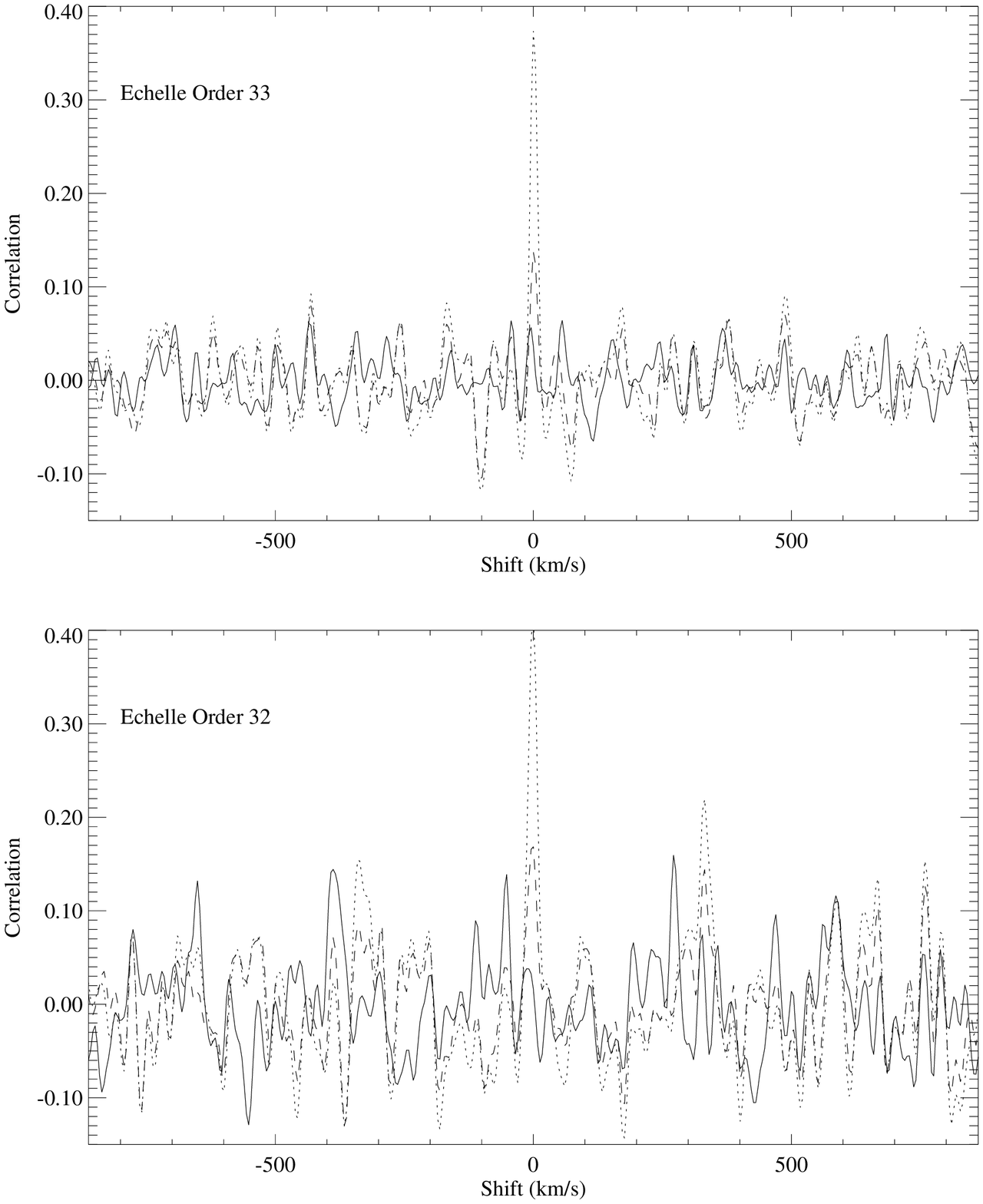}
\end{figure}

\end{document}